\documentclass[prc,preprint,tightenlines,floatfix,%
preprintnumbers,amsmath,amssymb]{revtex4}

\usepackage{graphicx}
\usepackage{bm}

\newcommand{\mc}[1]{\multicolumn{#1}}   
\newcommand{\vect}[1]              
           {\mbox{\boldmath$#1$}}  

\begin{document}

\title{The influence of the symmetry energy on the giant monopole
resonance of neutron-rich nuclei analyzed in Thomas-Fermi theory}

\author{M. Centelles$^1$}
\author{S. K. Patra$^2$}
\author{X. Roca-Maza$^1$}
\author{B. K. Sharma$^3$}
\author{P. D. Stevenson$^4$}
\author{X. Vi\~nas$^1$}

\affiliation{
$^1$Departament d'Estructura i Constituents de la Mat\`eria
and Institut de Ci\`encies del Cosmos,
Facultat de F\'{\i}sica, Universitat de Barcelona,
Diagonal {\sl 647}, {\sl 08028} Barcelona, Spain \\
$^2$Institute of Physics, Bhubaneswar {\sl 751 005}, India \\
$^3$Department of Nuclear and Atomic Physics,
Tata Institute of Fundamental Research, 
Homi Bhabha Road, Mumbai {\sl 400 005}, India \\
$^4$Department of Physics, University of Surrey, 
Guildford, Surrey, {\sl GU2 7XH}, UK}

\date{\today}


\begin{abstract}
We analyze the influence of the density dependence of the symmetry
energy on the average excitation energy of the isoscalar giant
monopole resonance (GMR) in stable and exotic neutron-rich nuclei by
applying the relativistic extended Thomas-Fermi method in scaling and
constrained calculations. For the effective nuclear interaction, we
employ the relativistic mean field model supplemented by an
isoscalar-isovector meson coupling that allows one to modify the
density dependence of the symmetry energy without compromising the
success of the model for binding energies and charge radii. The
semiclassical estimates of the average energy of the GMR are known to
be in good agreement with the results obtained in full RPA
calculations. The present analysis is performed along the Pb and Zr
isotopic chains. In the scaling calculations, the excitation
energy is larger when the symmetry energy is softer. The same
happens in the constrained calculations for nuclei with small and
moderate neutron excess. However, for nuclei of large isospin the
constrained excitation energy becomes smaller in models having a soft
symmetry energy. This effect is mainly due to the presence of
loosely-bound outer neutrons in these isotopes. A sharp increase of the
estimated width of the resonance is found in largely neutron-rich
isotopes, even for heavy nuclei, which is enhanced when the symmetry
energy of the model is soft. The results indicate that at large
neutron numbers the structure of the low-energy region of the GMR
strength distribution changes considerably with the density dependence
of the nuclear symmetry energy, which may be worthy of further
characterization in RPA calculations of the response function.
\end{abstract}

\maketitle

\vspace*{-14mm}

\section{Introduction}

The departure of the equation of state (EOS) of neutron-rich matter
($N\!>\!Z$) from the symmetric limit ($N\!=\!Z$) is governed by the
nuclear symmetry energy. This makes the knowledge of the symmetry
energy important for a variety of areas of nuclear physics (masses,
neutron densities, heavy ion collisions, etc.) and of astrophysics
(supernovae, neutron stars, nucleosynthesis, etc.)
\cite{ste05,li08,lat04,klahn06,xu09,bednarek09}. However, the
available experimental data on nuclei, which correspond mostly to
isotopes with small and moderate neutron-proton asymmetries, do not
constrain very precisely the value of the symmetry energy at the
saturation density. Rather, it seems to be better constrained at a
lower density around 0.10~fm$^{-3}$
\cite{brow00,horo1a,horo01b,todd03}, corresponding to an
average between the bulk and surface symmetry contributions in the
semi-empirical mass formula \cite{dan03,cen09,warda09}. All in all,
the density dependence of the symmetry energy is a basic nuclear
property that still is not sufficiently well known \cite{ste05,li08}.
This situation has prompted a recent ongoing effort to characterize
more closely the symmetry energy in the regime of subsaturation
densities, relevant for finite nuclei, from the phenomenology of
different nuclear observables
\cite{ste05,bar05,she07,tri08,kli07,li08,tsang09,dan03,cen09,warda09,roy09}. 
A proper understanding of the symmetry energy is particularly desirable
in order to predict ground-state and excitation properties of atomic
nuclei far from the naturally-occurring region of stability
\cite{oza01,khoa96,khoa04,nuc04,ENAM}.

The study of collective excitations of nuclei near and away from the
stability line is a salient source of information for nuclear
structure \cite{harakeh01,harakeh02,young99,gar07,monrozeau08}. For
instance, the isoscalar giant monopole resonance (GMR) constrains the
compression modulus $K_0$ of the EOS of symmetric nuclear matter. This
resonance has been accurately measured in heavy and medium mass nuclei
through recently improved $\alpha$-particle scattering experiments
\cite{young99,gar07}. By comparing the 1p-1h RPA results predicted by
Gogny \cite{blai95}, Skyrme \cite{farine97,hama97,colo04} and
relativistic mean field (RMF) interactions \cite{zhong01,todd05} with
the excitation energy measured in $^{208}$Pb ($14.17\pm0.28$ MeV
\cite{young99}), it is concluded that the value of the nuclear
incompressibility $K_0$ must lie in the range of 200--280 MeV.
Performing measurements of the GMR in exotic nuclei may deliver other
valuable informations and enhance our knowledge of the physics of
isospin-rich and loosely-bound nuclear systems
\cite{harakeh02,hama97,hama97b,monrozeau08,colo09}. However, it
represents an important experimental challenge due to e.g.\ the low
intensities in the radioactive ion beams. With the advent of new
technologies it is hoped that measurements of the breathing mode in
short-lived unstable nuclei will become feasible in the future
\cite{harakeh02}. Note that promising first successes have already
been demonstrated \cite{monrozeau08}. Although the value of the
excitation energy of the GMR is driven to a large extent by the
compression modulus of symmetric matter, it is known to be influenced
also by the density dependence of the symmetry energy in a
neutron-rich nucleus such as $^{208}$Pb
\cite{piek02,piek04,vret02,agra03,colo04,piek07b}. This influence is
expected to be amplified in isotopes with large neutron numbers, but
how and to which extent has not actually been assessed.

Our purpose in this paper is to investigate the effects of the density
dependence of the symmetry energy on the average excitation energy of
the GMR in isospin-rich nuclei, of heavy and medium mass, when moving
away from the stability valley. We ought to mention, however, that our
aim here is not pursuing to constrain the nuclear EOS from the present
calculations, but rather to analyze the effects that arise from the
existing uncertainty in the knowledge of the density dependence of the
symmetry energy. We advance that the calculations will show that the
impact on the excitation energy of the GMR is somehow moderate for
isotopes not far from stability, while the more notable effects are
found for nuclei with large neutron numbers, thereby (unfortunately)
rendering them more difficult to see in experiment. To describe the
effective nuclear interaction, we found convenient to use the model of
\cite{horo1a,horo01b,todd03} that consists of a standard RMF
Lagrangian plus an isoscalar-isovector nonlinear interaction between
the $\omega$ and $\rho$ meson fields. This coupling provides an
efficient way to change in a controlled manner the density dependence
of the symmetry energy without modifying the EOS of symmetric matter.
In particular, the compression modulus $K_0$ is not modified. When the
model is applied to finite nuclei, the variation of the strength of
the isoscalar-isovector coupling changes the properties of the neutron
density distribution (for example, the neutron rms radius), which have
large experimental uncertainties, but the predicted ground-state
binding energies and charge rms radii, which are accurately known in
experiment, remain basically unaltered. Thus, this RMF model allows
one to modify the density content of the symmetry energy while
preserving the agreement with the accurate ground-state information on
which the model has been calibrated previously (see
\cite{horo1a,horo01b,todd03,piek04,sil05}).

The consistent microscopic framework for the calculation of the
properties of the GMR is furnished by the Hartree-Fock plus RPA
theory, which corresponds to the small-amplitude limit of the
time-dependent Hartree-Fock method. The generalization of this theory
to deal with pairing in open-shell nuclei is known as quasiparticle
RPA (QRPA). The QRPA provides the full information on the response
function and centroid energies of the GMR\@. These microscopic
calculations are manageable but rather demanding if they are fully
self-consistent, which is accentuated when they are performed using
the canonical basis or in isotopes with large neutron numbers as it
has been recently discussed in \cite{colo09}. We will carry out
our analysis by studying the estimates of the average excitation
energy of the GMR that can be obtained without performing explicit RPA
calculations but which carry meaningful physical information on the
RPA strength. The typical example of this procedure is the sum-rule
approach to the RPA developed in the nonrelativistic framework using
Skyrme forces \cite{bohigas79,lipparini89,gleissl90}. The sum-rule
techniques provide average properties of the GMR such as the centroid
energy and the resonance width and have been successfully applied to
study the GMR in different nuclei and isotopic chains
\cite{blai95,bohigas79,lipparini89,gleissl90,cent05,sil06,colo04,colo09}. 
In the sum-rule approach one obtains the cubic and inverse
energy-weighted moments of the RPA strength function by means of,
respectively, scaling and constrained calculations of the uncorrelated
Hartree-Fock nuclear ground state. Note that these ground-state
scaling and constrained values are the {\em exact\/} RPA moments
\cite{bohigas79,lipparini89,gleissl90}. Because highly-accurate
ground-state calculations are simpler, the sum-rule techniques also
have been used in the literature as a means to check the
self-consistency of explicit RPA calculations of the strength function
\cite{hama97,cent05,sil06,colo09}. 

The validity of the sum-rule theorems when pairing is active (QRPA)
has been proven in recent publications \cite{khan02,colo09}. Capelli
et al.\ \cite{colo09} have made a careful study of constrained
Hartree-Fock (CHF), constrained Hartree-Fock-Bogoliubov (CHFB) and
QRPA calculations of the excitation energy of the GMR in the light
isotopic chains of Ca and Ni. The calculations have reached up to the
theoretical neutron drip line of the Skyrme SkM* force, i.e., up to
the isotopes $^{76}$Ca and $^{84}$Ni. Admitting that these isotopic
chains are known to terminate earlier in experiment than in the
predictions of SkM* and of most of the nuclear effective forces, and
knowing also that it will be difficult to measure the GMR in even less
exotic isotopes, the calculations spanning the whole theoretical
isotopic chains are very useful to test the predictions of nuclear
forces under extreme isospin conditions (which can take place in
astrophysical scenarios \cite{sil02}), as well as to test the
different many-body frameworks employed to deal with the problem. It
may be noted also that Khan \cite{Khan09b} has recently concluded that
knowing the GMR in a single isotope such as $^{208}$Pb may not suffice
to extract $K_0$ of nuclear matter and that the whole isotopic chain
should be measured in order to have a grasp on the various effects
acting on the GMR \cite{Khan09b}.

From the results of \cite{colo09} for Ca and Ni, it is seen that the
inclusion of pairing (CHFB and QRPA calculations) does not modify
significantly the behaviour shown by the GMR excitation energies
obtained in the CHF calculations along the isotopic chains. In the
case of the Ni chain, the atomic number not being as small as in Ca,
the CHFB and CHF values agree fairly well for most isotopes from
$^{56}$Ni up to the neutron drip line included, as noticed in figure~5
of \cite{colo09}. Of course, superfluidity is to be taken into account
if one wants to describe in detail the excitation spectra or if
precise values of the excitation energy within a few hundred keV in
specific nuclei are needed for comparison with accurate measurements
\cite{colo09,Khan09,Khan09b}. It has been pointed out in \cite{colo09}
that the QRPA calculations for isotopes having large neutron numbers
become much time consuming and that ensuring proper numerics in such
isospin-rich nuclei becomes somewhat difficult due to the weakly bound
neutrons. Indeed, the authors of \cite{colo09} state that in the
case of large neutron numbers very accurate QRPA calculations may
become almost prohibitive in keeping with the size of the basis.

The GMR is a collective excitation whose average properties in general
vary smoothly with the mass number $A$ and are rather insensitive to
shell effects, at least for not too light isotopes. Thus, a
description of the average properties of this resonance through
semiclassical approaches like the extended Thomas-Fermi (ETF) method
can be, for some purposes, a practical and interesting alternative. It
has been applied in different studies in the nonrelativistic
Skyrme-Hartree-Fock framework
\cite{gleissl90,cent90,liguo90,liguo91,cent05}. The results have shown
that the average excitation energies of the GMR obtained from scaling
and constrained quantal HF calculations are nicely described by the
corresponding semiclassical ETF counterparts. The ETF predictions of
GMR excitation energies become increasingly closer to the quantal
predictions with increasing atomic number, as it can be expected in
semiclassical methods. Indeed, in the calculations along the Pb
isotopic chain with the Skyrme force shown in figure~12 of
\cite{cent05}, the ETF scaling and constrained excitation energies
differ from the corresponding quantal values by less than 0.4~MeV and
they reproduce quite accurately the change of the quantal results in
going from the proton to the neutron drip line. Thus, from the
experience in nonrelativistic calculations, we believe that the
semiclassical scaling and constrained estimates of the average
excitation energy will be a useful and realistic procedure to have a
quick representation of the general trends of the GMR along isotopic
chains also in the relativistic mean field framework. And in heavy
systems such as the Pb isotopes, we expect that the semiclassical
calculations of the GMR average excitation energy will yield reliable
predictions practically on the quantitative level. Certainly, for a
precise microscopic description, the more demanding QRPA calculations
would have to be performed.

The semiclassical relativistic extended Thomas-Fermi (RETF) approach
\cite{centelles93a,centelles92,speicher93,centelles98,centelles93b} has
recently been applied in scaling and constrained calculations of the
GMR energy in stable isotopes in \cite{patra01,patra02}. The results
turn out to be in good agreement with the values extracted from full
relativistic RPA calculations \cite{patra01,patra02}. Let us mention
as a side comment that the RETF scaling procedure has been helpful in
obtaining the virial theorem for the relativistic mean field model
\cite{patra01} and that it has allowed self-consistent calculations of
the surface contribution $K_{\rm surf}$ to the nuclear
incompressibility in the relativistic theory \cite{patra02b}. In this
paper we will use the relativistic nuclear mean field model with an
isoscalar-isovector coupling mentioned previously together with the
RETF approach for describing the average excitation energy of the GMR
in the Pb and Zr isotopic chains, through scaling and constrained
calculations. The present results are expected to provide a realistic
indication on how the density dependence of the symmetry energy
impacts on the behaviour of the excitation energies and widths of the
GMR along isotopic chains of heavy and medium mass. In the next
section the basic theory is presented, in the third section the
results are discussed and the summary and conclusions are laid in the
fourth section. Some details about the semiclassical RETF calculations
are given in the appendices.

\section{Formalism}
\label{formal}

A detailed description of the scaling and constrained calculations of
the excitation energy of the GMR in the relativistic framework using
the RETF approach has been presented elsewhere \cite{patra01,patra02}.
Here, we shall restrict ourselves to outline the more relevant
expressions, incorporating the new contributions that arise from the
mixed $\omega$-$\rho$ meson interaction.

\subsection{ETF energy density of the relativistic mean field model}

The basic model for the present study is the relativistic mean field
theory of nucleons and mesons. As in the nonrelativistic problem, the
theory can be formulated in the Thomas-Fermi approach. When
corrections of order $\hbar^2$ (in gradients of the nucleon densities
and of the Dirac effective mass) to the pure Thomas-Fermi contribution
are included in the energy density, it is known as the relativistic
extended Thomas-Fermi approach
\cite{centelles93a,centelles92,speicher93,centelles98,centelles93b}.

The local energy density that we employ in the present work can be
written as follows:
\begin{eqnarray}
{\cal H} & = & {\cal E} + W \rho + B \rho_3 + {\cal A} \rho_{p} + 
\frac{1}{2 g_s^2} \left[ (\vect{\nabla}\Phi)^2 + m_{s}^2 \Phi^2 \right]
+ \frac{\kappa}{3!} \Phi^3 + \frac{\lambda}{4!} \Phi^4
\nonumber \\[3mm]
 & & \mbox{}
- \frac{1}{2 g_v^2} \left[ (\vect{\nabla} W)^2 + m_{v}^2 W^2 \right]
 - \frac{1}{2 g_{\rho}^2} \left[ (\vect{\nabla} B)^2 + m_\rho^2 B^2 
\right] - \Lambda_V B^2 W^2
 - \frac{1}{2} \left(\vect{\nabla} {\cal A} \right)^2 ,
 \label{eq0a}
\end{eqnarray}
where $\rho= \rho_{p}+\rho_{n}$ is the baryon density, $\rho_3=
{\textstyle\frac{1}{2}} (\rho_{p}-\rho_{n})$ is the isovector density,
and the term ${\cal E}$ will be described immediately below. The
energy density (\ref{eq0a}) contains scalar $\Phi= g_s
\phi_0(\vect{r})$, vector $W= g_v V_0(\vect{r})$ and vector-isovector
$B= g_{\rho} b_0(\vect{r})$ meson fields, as well as the Coulomb field
${\cal A}= e A_0(\vect{r})$. These fields represent, respectively,
$\sigma$-meson, $\omega$-meson, $\rho$-meson and photon exchange. The
cubic $\kappa\Phi^3$ and quartic $\lambda\Phi^4$ self-interactions of
the scalar field soften the EOS of symmetric nuclear matter around the
saturation point. In addition, the model is supplemented by a
nonlinear mixed coupling between the $\omega$-meson and the
$\rho$-meson fields. This $\Lambda_V B^2 W^2$ interaction was
introduced in \cite{horo1a} to soften the density dependence of
the symmetry energy and of the EOS of neutron matter, which are two
quantities that are predicted to be stiff by the majority of
relativistic mean field models.

In the quantum approach, the contribution ${\cal E}$ to the RMF energy
density in equation~(\ref{eq0a}) is given by ${\cal E} = \sum_\nu
\varphi_\nu^{\dagger} [ -i\,\vect{\alpha} \cdot\! \vect{\nabla} +
\gamma_0 (m-\Phi) ] \varphi_\nu$. Here, the $\varphi_\nu$ are
single-particle Dirac wave functions, $\vect{\alpha}$ and $\gamma_0$
are the usual Dirac matrices and $m$ denotes the rest mass of the
nucleons. The RETF representation of ${\cal E}$ consists of a pure
Thomas-Fermi part ${\cal E}_0$ plus a part ${\cal E}_2$, which is of
order $\hbar^2$ compared to ${\cal E}_0$. The nucleon variables of the
RETF energy density are the neutron and proton densities ($\rho_{n}$
and $\rho_{p}$) instead of the single-particle wave functions
$\varphi_\nu$. Thereby, ${\cal E}$ is a functional of the nucleon
densities and of the Dirac effective mass $m^{*}=m - \Phi$ in the RETF
approach. The gradient corrections contained in the ${\cal E}_2$ term
provide an improved description of the nuclear surface with respect to
the Thomas-Fermi treatment with only the ${\cal E}_0$ contribution.
The reader can find the full RETF expression of the functional ${\cal
E}$ in appendix~A. We also provide there the variational
Euler-Lagrange equations derived from equation~(\ref{eq0a}) in the RETF
approach for the nucleon densities and for the meson and photon fields. 

The energy density $\cal H$ is to be integrated over space to compute
the total energy of the nucleus. By means of partial integrations, and
on account of the variational meson field equations given in
appendix~A, it is possible to write the relativistic energy density of
a finite nucleus in the following form:
\begin{equation}
{\cal H} = {\cal E} +\frac{1}{2}\Phi \rho^{\rm eff}_{s}
+ \frac{\kappa}{3!} \Phi^3+ \frac{\lambda}{4!} \Phi^4
+\frac{1}{2} W \rho + \frac{1}{2} B \rho_3 + \Lambda_V B^2 W^2
+\frac{1}{2} {\cal A} \rho_{p} \,.
\label{eq1}
\end{equation}
Here, we have introduced the effective scalar density $\rho^{\rm
eff}_{s}$ defined as 
\begin{equation}
\rho^{\rm eff}_{s} = \rho_{s}
- \frac{\kappa}{2!} \Phi^2 - \frac{\lambda}{3!} \Phi^3. 
\label{eq1a}
\end{equation}
The form given in equation~(\ref{eq1}) for ${\cal H}$ turns out to be more
practical to perform the scaling of the equations in order to compute
the excitation energy of the GMR in the scaling formalism.

\subsection{Energy of the GMR in the scaling approach}
\label{escal}

The excitation energy of the isoscalar giant monopole resonance of a
finite nucleus of mass number $A$ is customarily written as
\begin{equation}
 E_{M} = \sqrt{ \frac{A K_A}{B_{M}} } \,,
\label{eq11}
\end{equation}
where $K_A$ is called the compression modulus or incompressibility of
the finite nucleus, and $B_{M}$ is called the mass or inertia parameter
of the monopole oscillation. In the scaling approach, the
incompressibility $K_A$ is obtained from the restoring force of the
monopole vibration modeled through a scaling transformation of the
nucleon densities:
\begin{equation}
 \rho_{q \alpha} (\vect{r}) = \alpha^3 \rho_q(\alpha\vect{r}) .
\label{eq4}
\end{equation}
Here, $\alpha$ denotes the scaling parameter and $q=\rm n, p$ refers
to neutrons or protons. The meson fields do not scale as simple powers
of $\alpha$ because of the finite-range character of the meson
interactions; we collect in appendix~B the variational field equations
for the scaled meson fields of the RMF model with the mixed
$\omega$-$\rho$ interaction. Following \cite{patra01,patra02}, it is
helpful to write the scaled Dirac effective mass $m^*_\alpha(\vect{r})
= m - \Phi_\alpha(\vect{r})$ in the form
\begin{equation}
 m^*_\alpha(\vect{r}) \equiv \alpha \, {\tilde m}^* (\alpha\vect{r}) 
\label{eq6} 
\end{equation}
because the energy density $\cal E$ and the scalar density $\rho_s$ of
the RETF model scale, respectively, as
${\cal E}_\alpha(\vect{r}) =
\alpha^4 \, {\cal E} [ \rho_q (\alpha\vect{r}) ,
                         {\tilde m}^* ( \alpha \vect{r} ) ]
\equiv \alpha^4 \, {\tilde{\cal E}} ( \alpha \vect{r})$
and as
$\rho_{\rm s\alpha} (\vect{r}) =
\alpha^3\, \rho_{\rm s} [ \rho_q (\alpha\vect{r}),
                         {\tilde m}^* (\alpha\vect{r}) ]
\equiv \alpha^3\, {\tilde\rho}_{\rm s} (\alpha\vect{r})$.
The tilded quantities ${\tilde{\cal E}}$ and
${\tilde\rho}_{\rm s}$ have the same expressions as ${\cal E}$ and
$\rho_{\rm s}$ given in appendix~A if one replaces $m^*$ by ${\tilde
m}^*$ in them. Taking into account the above, the scaled
form of the energy density of the present model becomes
\begin{equation}
{\cal H}_\alpha(\vect{r})
 =  \alpha^3 \bigg[ \alpha {\tilde{\cal E}}
+\frac{1}{2} \Phi_\alpha {\tilde\rho}^{\rm eff}_{s}
+\frac{1}{3!} \frac{\kappa}{\alpha^3}\Phi_\alpha^3
+\frac{1}{4!} \frac{\lambda}{\alpha^3} \Phi_\alpha^4
 +\frac{1}{2} W_\alpha \rho
+\frac{1}{2} B_\alpha \rho_3
+\frac{\Lambda_V}{\alpha^3} B_\alpha^2 W_\alpha^2
+\frac{1}{2} {\cal A}_\alpha\rho_{p} \bigg],
\label{eq7}
\end{equation}
with the definition
$\displaystyle
{\tilde\rho}^{\rm eff}_{s} =  {\tilde\rho}_{s}
- \frac{\kappa}{2!}  \frac{\Phi_\alpha^2}{\alpha^3}
- \frac{\lambda}{3!} \frac{\Phi_\alpha^3}{\alpha^3}$.
All densities and fields on the r.h.s.\ of (\ref{eq7}) depend on the
variable $\alpha \vect{r}$. 

The calculation of the first and second derivatives of the scaled
energy $E(\alpha)= \int d \vect{r} \, {\cal H}_\alpha(\vect{r})$ with
respect to $\alpha$ proceeds along the same lines described in
\cite{patra01,patra02}. The final result for the scaling
incompressibility is
\begin{eqnarray}
 K_A^{\rm scal} & = & \frac{1}{A}
 \left[ \frac{\partial^2}{\partial \alpha^2} \int
 d (\alpha \vect{r})
\frac{ {\cal H}_\alpha (\vect{r})}{\alpha^3} \right]_{\alpha=1} 
\nonumber \\[3mm]
& = & \frac{1}{A} \int d \vect{r} \left[
- m \frac{\partial {\tilde\rho}_{s}}{\partial \alpha} 
+ 3 \left( \frac{m_s^2}{g_s^2} \Phi^2 - \frac{m_v^2}{g_v^2} W^2
- \frac{m_\rho^2}{g_\rho} B^2 + \frac{\kappa}{3!} \Phi^3 \right) \right.
\nonumber \\[1.5mm]
& & \left. \mbox{}
- \left( 2 \frac{m_s^2}{g_s^2} \Phi + \frac{\kappa}{2} \Phi^2 \right)
\frac{\partial \Phi_\alpha}{\partial\alpha}
+ 2 \frac{m_v^2}{g_v^2} W
\frac{\partial W_\alpha}{\partial\alpha}
+ 2 \frac{m_\rho^2}{g_\rho^2} B
\frac{\partial B_\alpha}{\partial\alpha} \right]_{\alpha=1} ,
\label{eqFN30b}\end{eqnarray}
where $(\partial {\tilde\rho}_{s} / \partial
\alpha)_{\alpha=1} = - (\partial \rho_{s} / \partial m^*) \, [m^* +
(\partial \Phi_\alpha / \partial \alpha)_{\alpha=1}]$. 
It is interesting to note that all of the nucleonic contributions to
$K_A^{\rm scal}$ are summarized in the term with the scalar density
and that the massless photon field and the quartic meson couplings
such as $\lambda \Phi^4$ and $\Lambda_V B^2 W^2$ do not provide
explicit contributions. The expression (\ref{eqFN30b}) requires the
calculation of the derivatives of the meson fields with respect to the
scaling parameter $\alpha$ at equilibrium ($\alpha=1)$. These
derivatives can be computed by differentiation of the scaled meson
field equations with respect to $\alpha$ and by solving them at
$\alpha=1$, as we describe in appendix~B.

We will use the notation $E_{M}^{\rm scal}$ to refer to the excitation
energy of the GMR calculated in the scaling approach, namely,
\begin{equation}
E_{M}^{\rm scal} = \sqrt{\frac{A K^{\rm scal}_{A}}{B_{M}}} .
\label{eqEscal}
\end{equation}
The mass parameter of the monopole resonance is given by the
expression \cite{patra02,nishizaki87,zhu91}
\begin{equation}
B_{M} = \int d\vect{r} \,r^2 {\cal H} \,.
\label{eq13}
\end{equation}  
We note that in the nonrelativistic limit it reads
\begin{equation}  
B_{M}^{\rm nr} = \int d\vect{r} \,r^2 m \rho =
m\, A\, \langle r^2 \rangle ,
\label{eq13a} 
\end{equation} 
where $m$ is the nucleon rest mass. Because of the large contribution
of the nucleon rest mass term $m\rho$ to the energy density ${\cal H}$
in the integrand of equation~(\ref{eq13}), the value of the excitation
energy of the resonance is little modified by using either $B_{M}$ or
$B_{M}^{\rm nr}$ in the calculations.

\subsection{Energy of the GMR in the constrained approach}
\label{econs}

One also can study the average excitation energy of the giant monopole
resonance through constrained calculations
\cite{bohigas79,patra02,maru89,boer91,stoi94,stoi94b}. For that
purpose, one solves the problem associated with the constrained
functional 
\begin{equation}
 \int d \vect{r} [ {\cal H} - \eta\, r^2 \rho ] =
 E(\eta)  - \eta \int d \vect{r} r^2 \rho \,.
\label{eqFN56}
\end{equation}
The densities, fields and energy obtained from the solution of the
variational equations deduced from equation~(\ref{eqFN56}) depend on the
value of the parameter $\eta$ at which one performs the calculation.
By expanding $E(\eta)$ in a harmonic approximation about its minimum,
located at the ground-state rms radius $R_0$ (corresponding to
$\eta=0$), one obtains the compression modulus of the finite nucleus
in the constrained approach:
\begin{equation}
K^{\rm cons}_{A} = \frac{1}{A} R_0^2 \left(
\frac{\partial^2 E(\eta)}{\partial R_\eta^2}\right)_{\eta=0} .
\label{eqFN60}
\end{equation}
Here, $R_\eta^2=\langle r^2 \rangle_{\eta}$ is the square mass radius
of the nucleus computed with the density $\rho$ of
equation~(\ref{eqFN56}). One can show that equation~(\ref{eqFN60}) can
be rewritten in the following equivalent forms:
\begin{equation}
K^{\rm cons}_{A} = 
-4 R_0^4 \left(
  \frac{\partial R_\eta^2}{\partial \eta} \right)_{\eta=0}^{-1} =
4 A R_0^4 \left(
  \frac{\partial^2 E(\eta)}{\partial \eta^2} \right)_{\eta=0}^{-1} .
\label{eqFN60x}
\end{equation}
We have used (\ref{eqFN60x}) to check the numerical accuracy of our
calculations of $K^{\rm cons}_{A}$. Finally, the excitation energy of
the constrained isoscalar monopole vibration is computed as
\begin{equation}
E_{M}^{\rm cons} = \sqrt{\frac{A K^{\rm cons}_{A}}{B_{M}}} .
\label{eqEcons}
\end{equation}

\subsection{Comparison with the expressions in the nonrelativistic
sum-rule approach} 
\label{resemblance}

We would like to point out that there is a parallelism between the
scaling and constrained expressions for the GMR energy derived above
for the relativistic model and the known results in the
nonrelativistic formalism of sum rules. That is, one can define
suitable average excitation energies of the resonance such as
\begin{equation}
 E_3 \equiv \sqrt{\frac{m_3}{m_1}}
\qquad {\rm and} \qquad
 E_1 \equiv \sqrt{\frac{m_1}{m_{-1}}}
\label{eqFN60y}
\end{equation}
in terms of ratios of the integral moments $m_k = \int_0^\infty \!
d\omega\, \omega^k S(\omega)$ of the RPA strength function
$S(\omega)$. In the nonrelativistic RPA theory of the GMR
\cite{bohigas79,lipparini89,gleissl90}, it has been demonstrated that
the cubic energy-weighted moment $m_3= \int_0^\infty \! d\omega\,
\omega^3 S(\omega)$ of the RPA strength function is exactly equivalent
to the second derivative of the energy obtained from a monopole
scaling transformation. It is also known that the RPA energy-weighted
moment $m_1= \int_0^\infty \! d\omega\, \omega S(\omega)$ is
equivalent to
\begin{equation}  
m_1= \frac{2}{m} A \langle r^2 \rangle 
\label{eq13b} 
\end{equation} 
(``Thouless theorem''), and that the exact RPA inverse energy-weighted
moment $m_{-1}= \int_0^\infty \! d\omega\, S(\omega)/\omega$ can be
obtained from a constrained calculation of the Hartree-Fock ground
state: 
\begin{equation}
 m_{-1}= - \frac{1}{2} A \left(
 \frac{\partial R_{\eta}^2}{\partial \eta} \right)_{\eta=0} 
\label{eqsumr2} 
\end{equation}
(``dielectric theorem''). Taking into account these results and the
fact that $m_1= (2/m^2)\, B_{M}^{\rm nr}$, one sees an interesting
analogy between $E_3$ and $E_1$ of the nonrelativistic RPA formalism
of sum rules and $E_{M}^{\rm scal}$ of equations~(\ref{eqEscal}) and
(\ref{eqFN30b}) and $E_{M}^{\rm cons}$ of equations~(\ref{eqEcons})
and (\ref{eqFN60x}) given previously for the GMR average excitation
energy in the relativistic model, though to our knowledge the
sum-rule approach has not been derived in the relativistic RPA theory.

It is worth mentioning that the integral moments of the
nonrelativistic RPA strength function obey certain inequalities
\cite{bohigas79,lipparini89,gleissl90}. In particular, one has
\begin{equation} 
 E_1 \,\leq\, E_x= \frac{m_1}{m_0} \,\leq\, E_3 \,,
\label{centroid}
\end{equation}
where $E_x$ is the centroid energy of the GMR evaluated from the ratio
of the $m_1$ moment to the $m_0$ moment. That is, $E_1$ is a lower
limit for the mean energy (centroid) of the resonance, whereas $E_3$
is an upper limit. It is clear from the definition of the $m_3$ and
$m_{-1}$ moments that $E_3$ and $E_1$ explore different energy domains
of the collective excitation, i.e., $E_3$ is more influenced by the
high-energy region of the strength function $S(\omega)$, whereas $E_1$
is more sensitive to the low-energy components of $S(\omega)$. It can
be shown \cite{bohigas79} that the quantity
\begin{equation}
 \sigma = \frac{1}{2} \sqrt{E_3^2 - E_1^2}
\label{width}
\end{equation}
provides an upper bound for the width of the giant monopole resonance
in the RPA approximation. The calculations carried out for stable
nuclei show that $E_3$ and $E_1$ become increasingly close with
increasing nuclear mass. Therefore, in a stable heavy nucleus the
value of the width $\sigma$ is small and the monopole strength becomes
concentrated around a single peak \cite{bohigas79}, which is the
situation found in experiment. Following equation~(\ref{width}) and
the above comments, in the discussion of our results the spread
between the calculated $E_{M}^{\rm scal}$ and $E_{M}^{\rm cons}$
values will provide an indication about the width of the GMR.

\section{Results and discussions}

\subsection{Comparison of relativistic ETF and Hartree-RPA results}
\label{comparison}

First, in this section we test with some examples the results of the
relativistic extended Thomas-Fermi (RETF) approach vis-\'a-vis
relativistic Hartree and RPA calculations. To this end, we report in
table~1 semiclassical RETF results for the binding energy and the
excitation energy of the GMR in the nuclei $^{208}$Pb and $^{90}$Zr.
We also include in the comparison the value of the neutron skin
thickness $\Delta R_{np}= R_n-R_p$ (the difference between the neutron
and proton rms radii of the nucleus) because this quantity is very
sensitive to the density dependence of the symmetry energy
\cite{brow00,type01,cen09,bald04}. The results have been
computed with the NL3 mean field interaction \cite{lala97}
supplemented by the nonlinear $\Lambda_V$ coupling between the
$\omega$-meson and $\rho$-meson fields \cite{horo1a}. As pointed out
in the Introduction, this model, through the isoscalar-isovector meson
coupling, allows one to change the symmetry energy leaving unchanged
the properties of the EOS of symmetric nuclear matter. By modifying
the value of $\Lambda_V$ one can investigate the changes in finite
nuclei due to the variation of the density dependence of the symmetry
energy almost without affecting the previous quality of the model for
binding energies and charge radii 
\cite{horo1a,horo01b,todd03,piek04,sil05}.

\begin{table}
\caption{Relativistic Hartree-RPA and relativistic extended
Thomas-Fermi values obtained with the NL3 model plus the
isoscalar-isovector coupling $\Lambda_V$ for the binding energy per
particle (MeV), neutron skin thickness $\Delta R_{np}= R_n-R_p$ (fm),
and excitation energy of the GMR (MeV) in $^{208}$Pb and $^{90}$Zr.
The GMR centroid energies $m_1/m_0$ are taken from \cite{piek04}.} 
\begin{ruledtabular}
\begin{tabular}{lccccccc}
$^{208}$Pb: & \\
 &\mc{3}{c}{quantal Hartree-RPA} &\mc{4}{c}{semiclassical RETF} \\
 \cline{2-4} \cline{5-8}
Model & $B/A$ & $\Delta R_{np}$ & $m_1/m_0$ &
$B/A$ & $\Delta R_{np}$ & $E_{M}^{\rm cons}$ & $E_{M}^{\rm scal}$ \\
%
%
\hline
NL3.00&7.87&0.28&14.32& 7.99 & 0.23&13.92& 14.58\\
NL3.01&7.89&0.25&14.43& 8.01 & 0.20&13.97& 14.61\\
NL3.02&7.91&0.22&14.57& 8.02 & 0.17&14.07& 14.70\\
NL3.03&7.91&0.20&14.74& 8.03 & 0.15&14.20& 14.83\\
\hline
$^{90}$Zr: & \\
 &\mc{3}{c}{quantal Hartree-RPA} &\mc{4}{c}{semiclassical RETF} \\
 \cline{2-4} \cline{5-8}
Model & $B/A$ & $\Delta R_{np}$ & $m_1/m_0$ &
$B/A$ & $\Delta R_{np}$ & $E_{M}^{\rm cons}$ & $E_{M}^{\rm scal}$ \\
\hline
NL3.00&8.69&0.11&18.62& 8.91 & 0.10&19.04& 19.53\\
NL3.01&8.69&0.10&18.67& 8.91 & 0.09&19.05& 19.54\\
NL3.02&8.70&0.09&18.69& 8.92 & 0.08&19.08& 19.58\\
NL3.03&8.70&0.08&18.75& 8.92 & 0.07&19.12& 19.62\\
\end{tabular}
\end{ruledtabular}
\end{table}

For the coupling constant $\Lambda_V$ we use the values 0.00, 0.01,
0.02, and 0.03 (models denoted as NL3.00, NL3.01, NL3.02 and NL3.03
in table~1). It is known that the binding energy of nuclei constrains
an average between the symmetry energy at saturation and the surface
symmetry energy, rather than the individual value of the symmetry
energy at saturation \cite{horo1a,horo01b,todd03,dan03,cen09}. Thus,
as done in earlier literature
\cite{horo1a,horo01b,todd03,piek04,sil05}, we constrain the symmetry
energy $S(\rho)$ of the models to take a value of 25.67 MeV at a Fermi
momentum $k_F=1.15$ fm$^{-1}$ (i.e., at a subsaturation density
$\rho\approx0.10$ fm$^{-3}$). This implies that the coupling constant
$g_{\rho}$ of the NL3 interaction has to be suitably adjusted for each
value of the isoscalar-isovector $\Lambda_V$ coupling. As a
consequence, the value $S(\rho_0)$ of the symmetry energy at the
saturation density $\rho_0$ changes with the $\Lambda_V$ parameter. We
find the results $S(\rho_0)= 37.4$, 35.0, 33.2 and 31.7 MeV when we
vary $\Lambda_V$ from 0.00 to 0.03. That is, the symmetry energy
$S(\rho)$ of the model around saturation becomes softer (it increases
more slowly with the nuclear density) when the value of the
$\Lambda_V$ parameter is larger. Further details on the behaviour of
$S(\rho)$ in the NL3 model supplemented by the $\Lambda_V$ coupling
can be found in \cite{todd03}. 

Table~I shows that by increasing the $\Lambda_V$ coupling the neutron
skin thickness of a neutron-rich nucleus is significantly reduced,
i.e., the models with a softer symmetry energy give rise to smaller
neutron skins. Here, this is achieved while maintaining the proton rms
radius of the nucleus essentially intact. For example, in the quantal
Hartree calculation with the NL3.00 model we find $R_p= 5.466$ fm for
$^{208}$Pb and 4.199 fm for $^{90}$Zr, while the respective values
using NL3.03 are 5.478 fm and 4.210 fm. The changes in $R_p$ are thus
less than 0.5\%. Moreover, one sees in Table~I that also the binding
energies are little altered by $\Lambda_V$. Both $^{208}$Pb and
$^{90}$Zr are slightly more bound at large $\Lambda_V$ but the changes
in $B/A$ are again no larger than 0.5\%. Thus, the predictions for
charge radii and binding energies of the original NL3 model are not
seriously compromised by the additional $\Lambda_V$ coupling, which
otherwise largely softens the symmetry energy.

From table~1 we observe that in the semiclassical RETF approach the
binding energies of $^{208}$Pb and $^{90}$Zr are larger and the
neutron skins smaller, than in the corresponding quantal Hartree
calculations. In principle, this fact is not to be considered a lack
of accuracy of the Thomas-Fermi-like methods because semiclassical and
quantal calculations must differ by shell effects. It is well known
that the semiclassical ground-state calculations by construction
smooth out the quantal shell effects and provide the average part of
the quantal binding energy \cite{RS80,bra85,Bra97}.
Actually, in the semiclassical calculations one works with the
particle densities and completely avoids to calculate single-particle
wave functions. By this same reason, the semiclassical approach yields
smooth average densities and energies that do not show the quantum
shell oscillations. The nuclear ETF functionals with gradient
corrections up to order $\hbar^2$ always produce some overbinding
(between a $\sim$2\% in heavy nuclei and a $\sim$10\% in light nuclei)
and yield smaller rms radii (between $\sim$1\% and $\sim$4\%) than the
quantal mean field calculations
\cite{bra85,cent90,centelles93a,centelles92}. The quantal corrections,
which are more prominent in magic nuclei (our situation in table~1),
can eventually be added perturbatively to the semiclassical binding
energies using techniques such as the Strutinsky shell-correction
method \cite{Bra97} or the expectation value method
\cite{bra85,centelles93a,centelles92}.

\begin{figure}[t]
\includegraphics[width=0.70\columnwidth,angle=0,clip=true]
{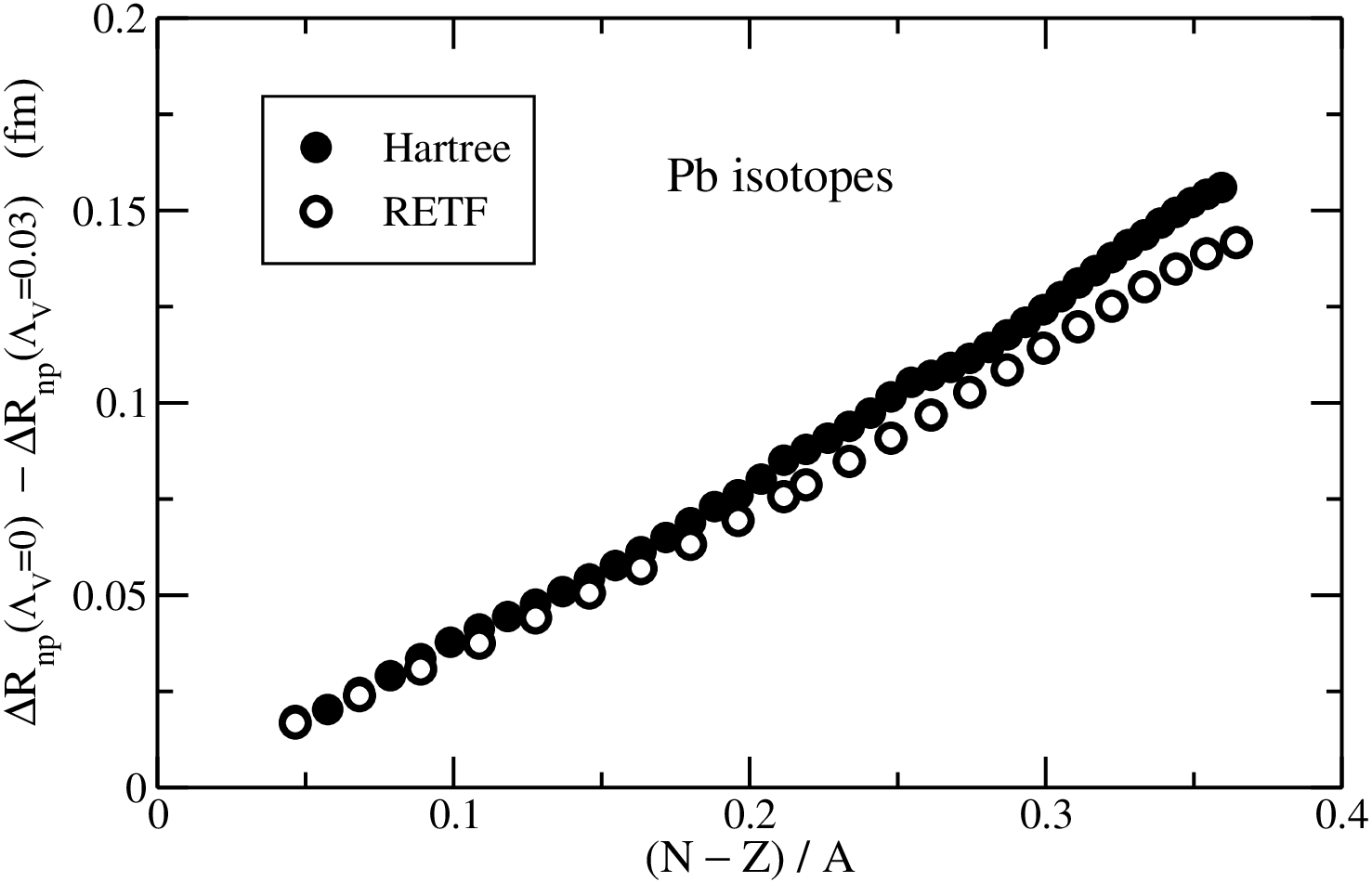}
\caption{Difference between the values of the neutron skin thickness
$\Delta R_{np}$ obtained with $\Lambda_V=0$ and with $\Lambda_V=0.03$
for lead isotopes. The results of the RETF method are compared
with the results of quantal Hartree calculations.}
\label{f_skindif}
\end{figure}

Our focus here is somehow more about the change of the results with
the softness of the symmetry energy than on the absolute values
themselves. In this respect, one has to note that the quantal and
semiclassical results reported in table~1 show both practically the
same variation as a function of the $\Lambda_V$ coupling. The binding
energies remain similarly constant and the neutron skins decrease by
almost the same amount of 0.08~fm in $^{208}$Pb and of 0.03~fm in
$^{90}$Zr in both the Hartree and RETF calculations when $\Lambda_V$
is changed from 0.00 to 0.03. Therefore, the RETF approach predicts
practically the same variation with $\Lambda_V$ as the quantal Hartree
calculations. This implies that the rate of change of the results with
the softness of the symmetry energy in the present model is well
described by the RETF method. As a further test, which is motivated
also because later we will study RETF results along isotopic chains,
in figure~\ref{f_skindif} we plot for Pb isotopes the difference between
the value of the neutron skin thickness of the nucleus calculated with
$\Lambda_V=0.00$ and with $\Lambda_V=0.03$, as a function of the
parameter $I=(N-Z)/A$. We observe by comparison with the corresponding
quantal Hartree results, that the RETF approach performs quite
satisfactorily along the Pb isotopic chain in predicting the right
size of the modification induced by the change of the symmetry energy
of the model.

With regard to the average excitation energies of the GMR, table~1
shows the scaled and constrained values (i.e., $E_{M}^{\rm scal}$ of
section \ref{escal} and $E_{M}^{\rm cons}$ of Section \ref{econs}) for
$^{208}$Pb and $^{90}$Zr calculated in the RETF approach. We compare
our RETF results with the values of the centroid energy $m_1/m_0$ of
the GMR extracted from the relativistic RPA (RRPA) strength, which
have been reported in \cite{piek04}. We see that for the
$^{208}$Pb nucleus, the RRPA centroid energies very nicely lie in
between the RETF constrained and scaled estimates, as it is required
in the nonrelativistic sum-rule approach to the RPA
\cite{bohigas79,lipparini89,gleissl90} (recall equation~(\ref{centroid})
of section \ref{resemblance}). In the case of a lower-mass nucleus
such as $^{90}$Zr, the semiclassical excitation energies are not
expected to be as successful as for the heavier $^{208}$Pb. Indeed, we
see in table~1 that the RETF constrained and scaled estimates in
$^{90}$Zr overestimate a little the RRPA centroid energies. The
discrepancies are nevertheless moderate, with differences no larger
than 2\% for $E_{M}^{\rm cons}$ and 5\% for $E_{M}^{\rm scal}$ with
respect to the RRPA $m_1/m_0$ value of $^{90}$Zr. And, as it happens
in the case of $^{208}$Pb, the semiclassical average excitation
energies of $^{90}$Zr as a function of the $\Lambda_V$ coupling
display the tendency of the RRPA excitation energies. In summary, we
can conclude from the results discussed in the present section that
the RETF approach is a reasonable starting point to study the
influence of the density dependence of the symmetry energy on the GMR
of neutron-rich nuclei.

\subsection{GMR in heavy neutron-rich nuclei}
\label{gmrheavy}

As mentioned in the Introduction, we found in \cite{cent05}
within the Skyrme-Hartree-Fock framework that the semiclassical
calculations are able to describe realistically the tendencies of the
excitation energy of the GMR in regions away from the stability
valley, especially for heavy nuclei such as the lead isotopes where a
semiclassical approach performs better. In the present section we
analyze with the RETF approach to the relativistic mean field model
the effects of the density dependence of the symmetry energy on the
excitation energy of the GMR in heavy neutron-rich nuclei by
contrasting the results for the nucleus $^{266}$Pb, which we take as
an example of a heavy system with a large neutron excess away from
stability, with the results for the stable isotope $^{208}$Pb.

\begin{figure}[t]
\includegraphics[width=0.90\columnwidth,angle=0,clip=true]
{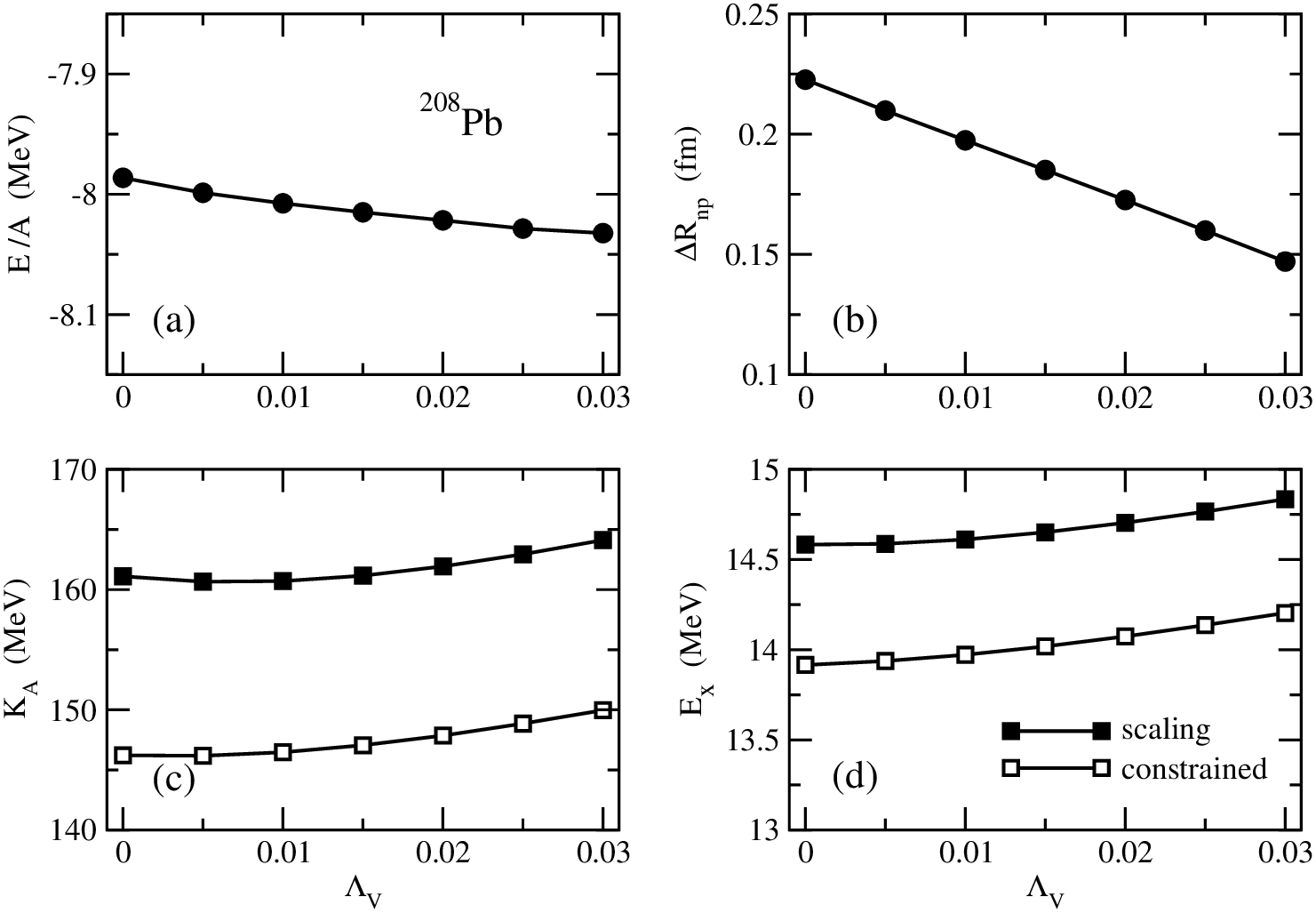}
\caption{Dependence with the value of the
isoscalar-isovector coupling $\Lambda_V$ of the energy per particle
(a), neutron skin thickness (b), finite nucleus compression modulus
(c) and average excitation energy of the GMR (d) for the $^{208}$Pb
nucleus.} 
\label{f_pb208}
\end{figure}

\begin{figure}[t]
\includegraphics[width=0.90\columnwidth,angle=0,clip=true]
{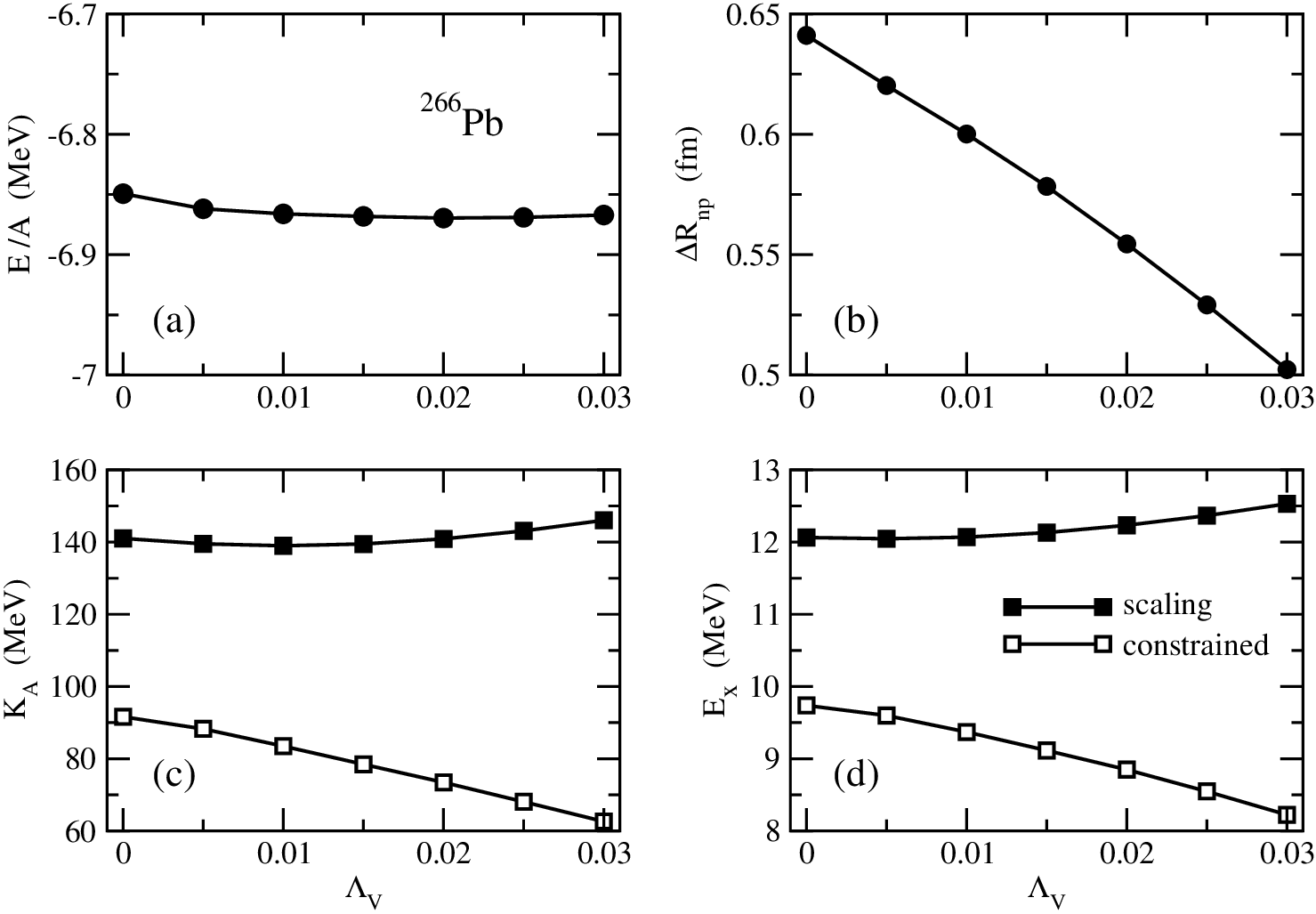}
\caption{Dependence with the value of the
isoscalar-isovector coupling $\Lambda_V$ of the energy per particle
(a), neutron skin thickness (b), finite nucleus compression modulus
(c) and average excitation energy of the GMR (d) for the $^{266}$Pb
nucleus.} 
\label{f_pb266}
\end{figure}

We display in figure~\ref{f_pb208} for $^{208}$Pb and in
figure~\ref{f_pb266} for $^{266}$Pb the calculated values of the
energy per particle $E/A$, the neutron skin thickness $\Delta R_{np}=
R_n-R_p$, the compression modulus of the finite nucleus $K_A$ and the
average excitation energy $E_x$ of the GMR as a function of the
coupling constant $\Lambda_V$ in the relativistic model. We recall
that the coupling $\Lambda_V$ induces a change of the density
dependence of the symmetry energy. The model has a stiff symmetry
energy at $\Lambda_V=0.00$, while it has a soft symmetry energy at
$\Lambda_V=0.03$. The RMF Lagrangian supplemented by the nonlinear
interaction between the $\omega$ and $\rho$ meson fields leaves the
binding energy of the nucleus almost independent of $\Lambda_V$, at
least for moderate values of this coupling. We note from the results
for $^{266}$Pb in figure~\ref{f_pb266} that this happens not only in
the case of stable nuclei, which we have seen previously for
$^{208}$Pb and $^{90}$Zr in table~1, but also in the case of much more
neutron-rich nuclei. The neutron thickness $\Delta R_{np}$ decreases
almost linearly with an increase of $\Lambda_V$ in both $^{208}$Pb and
$^{266}$Pb. The reduction of the $\Delta R_{np}$ value at
$\Lambda_V=0.03$ amounts to about a $30\%$ of the original neutron
skin thickness at $\Lambda_V=0.00$. The proton rms radius $R_p$ of
$^{208}$Pb and $^{266}$Pb has been found to remain nearly constant
with $\Lambda_V$, and therefore the decrease of $\Delta R_{np}$ is due
almost exclusively to the decrease of $R_n$. It is worth mentioning
that the alluded variation of the neutron skin thickness of $^{208}$Pb
in the present calculations is close to roughly a 1\% change in the
neutron rms radius $R_n$ of this nucleus. That is, the range of
variation we have allowed in the density dependence of the symmetry
energy by changing $\Lambda_V$ between 0.00 and 0.03 induces a change
in $R_n$ of $^{208}$Pb around 1\%. In this sense, it is consistent
with the uncertainty that will remain in this observable after the
PREX experiment at the Jefferson Laboratory \cite{mich03}, which aims
to measure $R_n$ of $^{208}$Pb to 1\% accuracy via parity-violating
electron scattering.

The average excitation energy of the GMR calculated in the RETF
approach involves the knowledge of the finite nucleus compression
modulus $K_A$, or restoring force of the resonance. We display $K_A$
in figures~\ref{f_pb208}(c) and \ref{f_pb266}(c) for $^{208}$Pb and
$^{266}$Pb, respectively. Let us discuss first the change of $K_A$
with the mass number $A$ in isotopes. We see from the results for
$^{208}$Pb and $^{266}$Pb that the finite nucleus compression modulus
decreases when $A$ increases. This behaviour may be easily understood
from the so-called leptodermous expansion of $K_A$ \cite{blai80}:
\begin{equation}
K_A = K_0+K_{\rm surf}A^{-1/3}+K_\tau I^2+K_{\rm Coul}Z^2A^{-4/3}
+ \cdots \,,
\label{lepto}
\end{equation} 
where $K_0$ is the compression modulus of symmetric infinite nuclear
matter and $K_{\rm surf}$, $K_\tau$ and $K_{\rm Coul}$ are the leading
surface, volume-symmetry and Coulomb corrections. The coefficients
$K_{\rm surf}$, $K_\tau$, and $K_{\rm Coul}$ in the scaling model are
negative, and in particular $K_\tau$ may be large
\cite{blai80,chossy97,patra02b,colo04,sagawa07,piek09}. Within an
isotopic chain, the nuclear charge $Z$ is fixed and therefore the
neutron-proton asymmetry $I=(N-Z)/A$ increases when $A$ increases.
Then, although the surface and Coulomb contributions in (\ref{lepto})
become a little less negative with increasing $A$, the increase of the
negative volume-symmetry contribution $K_\tau I^2$ dominates.
Therefore, $K_A$ in isotopes decreases with larger $A$. We stress that
the expansion (\ref{lepto}) is provided here for indicative purposes
only and that in all results we have computed $K_A$ self-consistently.

The scaling value $K_A^{\rm scal}$ of the finite nucleus
incompressibility, given by equation~(\ref{eqFN30b}) of section
\ref{escal}, is displayed in figures~\ref{f_pb208}(c) and
\ref{f_pb266}(c) by the upper curves with solid symbols. For both
isotopes $^{208}$Pb and $^{266}$Pb, $K_A^{\rm scal}$ shows an upward
trend with larger $\Lambda_V$. This variation of $K_A^{\rm scal}$ with
$\Lambda_V$ is due to the softening of the density dependence of the
symmetry energy. It modifies the value of $K_\tau$ and consequently
$K_A$ (of course, $K_\tau$ does not describe alone the whole effect as
there are surface-symmetry and other corrections neglected in
(\ref{lepto}) \cite{patra02b}). The present models have all
$K_0=271.5$ MeV, while the $K_\tau$ coefficient changes from $-699$
MeV in NL3.00 to $-381$ MeV in NL3.03. The recent empirical
extractions of $K_\tau$ suggest $K_\tau$ values in this range. For
example, isospin diffusion in nuclear reactions favours the constraint
$K_\tau=-500\pm50$ MeV \cite{li08}, the breathing mode of Sn isotopes
suggests $K_\tau=-550\pm100$ MeV \cite{gar07} (or $-500\pm50$ MeV
\cite{sagawa07}) and neutron skins of nuclei point to
$K_\tau=-500^{+125}_{-100}$ MeV \cite{cen09}.

The excitation energy $E_{M}^{\rm scal}$ of the monopole oscillation
evaluated in the scaling approach is determined by the ratio of the
restoring force $K_A^{\rm scal}$ to the mass parameter $B_{M}$
(section \ref{formal}). As indicated, to replace $B_{M}$ by its
nonrelativistic limit $B_{M}^{\rm nr}$ is a very good approximation.
In practice, this implies that $B_{M}$ varies with the parameters of
the nuclear interaction essentially in the same way as $\langle r^2
\rangle$ of the nucleus does, see equation~(\ref{eq13a}). When
$\Lambda_V$ is increased, $\langle r^2 \rangle$ of the matter
distribution of the nucleus decreases because of the reduction of the
neutron rms radius originated by the softer symmetry energy. As a
consequence, the mass parameter of the GMR (analogously, the $m_1$ sum
rule (\ref{eq13b}) in the nonrelativistic language) also decreases.
This effect combined with the growing tendency of $K_A^{\rm scal}$
with $\Lambda_V$ implies that the scaling excitation energy of the GMR
will be enhanced with a softer symmetry energy, which is confirmed in
figures~\ref{f_pb208}(d) and \ref{f_pb266}(d) where $E_{M}^{\rm scal}$
is plotted. We recall that this enhancement of $E_{M}^{\rm scal}$ is
due to the different density dependence of the symmetry energy alone
because the incompressibility $K_0$ of symmetric matter is forced to
be the same in all the interactions considered here.

Let us now analyze the results obtained through the constrained
calculations described in section \ref{econs}. For the stable nucleus
$^{208}$Pb, we see in figure~\ref{f_pb208} that $K_A^{\rm cons}$ and
$E_{M}^{\rm cons}$ have a similar behaviour to the discussed scaling
results as a function of the density dependence of the symmetry
energy. The separation between $E_{M}^{\rm scal}$ and $E_{M}^{\rm
cons}$ is indicative of the width of the GMR resonance (see
equation~(\ref{width})), and therefore this quantity is basically
constant with the symmetry energy in $^{208}$Pb. In contrast to this
situation in the stable $^{208}$Pb nucleus, in the more neutron-rich
isotope $^{266}$Pb we see in figure~\ref{f_pb266} that the change of
$K_A^{\rm cons}$ and $E_{M}^{\rm cons}$ with the symmetry energy
exhibits a different pattern from the corresponding scaling results.
For $^{266}$Pb, both quantities $K_A^{\rm cons}$ and $E_{M}^{\rm
cons}$ not only do not increase with $\Lambda_V$, but they decrease
significantly.

To interpret the reduction of $E_{M}^{\rm cons}$ and $K_A^{\rm cons}$
with $\Lambda_V$ in $^{266}$Pb, it will be helpful to rewrite the
expression $E_{M}^{\rm cons} = (A K^{\rm cons}_{A}/B_{M})^{1/2}$ of
equation~(\ref{eqEcons}) by approximating $B_{M}$ with its
nonrelativistic form $B_{M}^{\rm nr}= m\, A\, \langle r^2 \rangle$,
and to write $K^{\rm cons}_A$ using the first expression given in
equation~(\ref{eqFN60x}). This results in
\begin{equation}
E_{M}^{\rm cons} \simeq \sqrt{ \frac{2 A \langle r^2 \rangle/m}
{-\frac{1}{2} A (\partial R_{\eta}^2 / \partial \eta)_{\eta=0}} } 
\equiv \sqrt{\frac{\overline{m}_1}{\overline{m}_{-1}}} \,,
\label{emcons2}
\end{equation}
which in terms of the defined quantities $\overline{m}_1$ and
$\overline{m}_{-1}$ has the same form that $E_{M}^{\rm cons}$ takes in
the nonrelativistic RPA formalism of sum rules (i.e., $E_1$ of
equation~(\ref{eqFN60y})) \cite{bohigas79,lipparini89,gleissl90}. It
is to be noted that the denominator of (\ref{emcons2}) is very
sensitive to the presence of loosely bound nucleons in the nucleus
\cite{cent05}, which is precisely the situation that occurs in
$^{266}$Pb. When one is approaching the neutron drip line, the
negative chemical potential $\mu_n$ of neutrons becomes close to
vanishing, implying that the system contains weakly bound neutrons.
These neutrons are very soft against compression and in the
constrained calculations they give rise to a large increase of the
absolute value of $(\partial R_{\eta}^2 / \partial \eta)_{\eta=0}$,
i.e., of $\overline{m}_{-1}$ in~(\ref{emcons2}). This effect has been
found in \cite{cent05} in constrained HF (and ETF) calculations for
nonrelativistic Skyrme forces, where it can be related to the
enhancement of the RPA strength in the low-energy region produced by
the increase of the transitions to the continuum at large neutron
numbers. The same effect is found in the recently published QRPA and
CHFB Skyrme calculations, which take account of the pairing
correlations, performed in the light isotopic chains of Ca and
Ni~\cite{colo09}. In our present calculations, the increase of
$\overline{m}_{-1}$ at large neutron numbers is accentuated the softer
is the symmetry energy (note that parameter sets having a softer
symmetry energy have also weaker binding energy of the last bound
neutrons \cite{todd03} and therefore reinforce the effect). This
explains the decreasing trend of $E_{M}^{\rm cons}$ with $\Lambda_V$
in $^{266}$Pb noticed in figure~\ref{f_pb266}. The decrease of
$K_A^{\rm cons}$ with $\Lambda_V$ seen in the same figure is similarly
understood by recalling equation~(\ref{eqFN60x}) that relates $K_A^{\rm 
cons}$ with $(\partial R_{\eta}^2 / \partial \eta)_{\eta=0}^{-1}$.

\subsection{GMR along the Pb and Zr isotopic chains}

In this section we would like to investigate at which values of the
neutron excess of the nucleus the effects coming from the symmetry
energy become more relevant in the excitation energies of the GMR in
isotopic chains of heavy (Pb) and medium (Zr) mass. Within the
Skyrme-Hartree-Fock framework, in \cite{cent05} we analyzed the
excitation energy and the resonance width of the GMR in some isotopic
chains with the sum-rule approach (i.e., through scaling and
constrained calculations). In the case of a small atomic number such
as the Ca isotopes, the semiclassical ETF results were able to
describe the global variation of the quantal HF excitation energies
but could not reproduce them quantitatively. However, the ETF results
for the GMR excitation energies become progressively more accurate
with larger atomic number. In the case of the Pb chain, the ETF
scaling and constrained excitation energies were found to be in very
good agreement with the quantal HF values on the quantitative level
from the proton to the neutron drip line (the differences between ETF
and quantal values not exceeding 0.4~MeV) \cite{cent05}. Thus, from
the experience in the nonrelativistic calculations, as well as
suggested by the comparison of RETF and RRPA results made in table~1,
the present semiclassical predictions for the GMR are expected to be
sufficiently realistic, and rather accurate quantitatively (compared
to quantal calculations) in the case of the Pb isotopes.

\begin{figure}[t]
\includegraphics[width=0.70\columnwidth,angle=0,clip=true]
{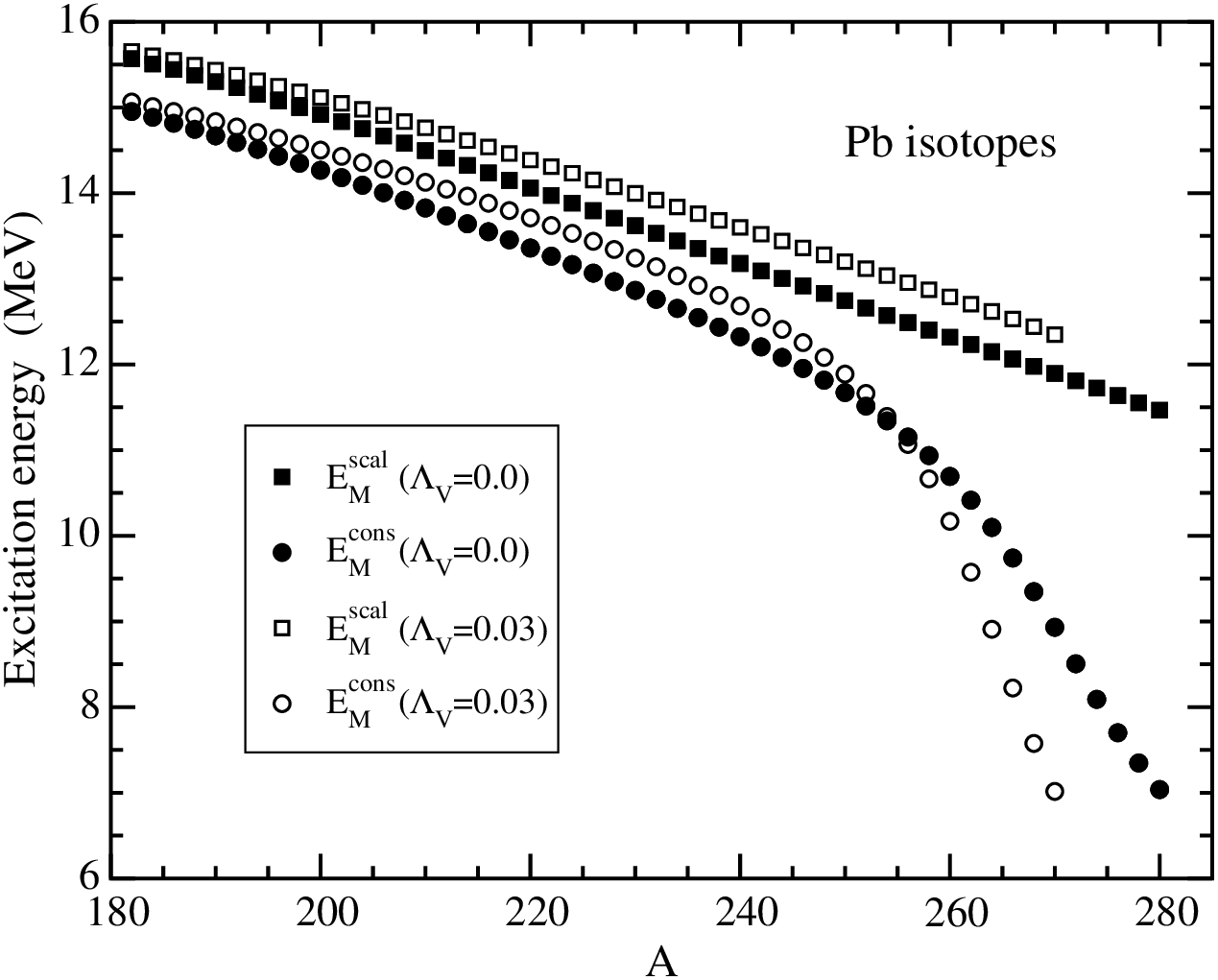}
\caption{Mass-number dependence of the excitation energy of the giant
monopole resonance in the Pb isotopic chain for the NL3 model with
$\Lambda_V=0$ (stiff symmetry energy) and $\Lambda_V=0.03$ (soft
symmetry energy), in the scaling and constrained approaches.}
\label{f_pb_ex}
\end{figure}

We depict in figure~\ref{f_pb_ex} the scaling and constrained average
excitation energies of the GMR for the Pb isotopic chain. Here we only
use the values 0.00 and 0.03 for the nonlinear $\omega$-$\rho$
interaction $\Lambda_V$, representing stiff and soft density
dependences of the symmetry energy. In the semiclassical calculations,
the vanishing of the neutron chemical potential $\mu_n$ defines the
location of the neutron drip line. We see in figure~\ref{f_pb_ex} that
the latter is reached at a smaller mass number when the model has a
soft symmetry energy, in agreement with the earlier observations of
\cite{todd03}. Note that shell effects are absent in these
semiclassical predictions of the drip line (a quantal Hartree
calculation with NL3 predicts the Pb neutron drip line at $A=266$).
The proton drip line occurs at neutron number $N$ not much larger than
$Z$. Therefore, because one deals with almost symmetric matter, the
position of the proton drip line is unaffected by the density
dependence of the symmetry energy.

The calculated GMR excitation energies show in figure~\ref{f_pb_ex} a
downward tendency with increasing mass number $A$ when one proceeds
along the isotopic chain, by similar reasons to those pointed out in
our previous discussion about the $^{208}$Pb and $^{266}$Pb isotopes.
The downward trend is more noticeable for $E_{M}^{\rm cons}$ than for
$E_{M}^{\rm scal}$ in the isotopes with larger neutron numbers. The
scaling excitation energies $E_{M}^{\rm scal}$ obtained using a
parameter set with a soft symmetry energy ($\Lambda_V=0.03$) are
systematically higher in the whole isotopic chain than if the symmetry
energy is stiff ($\Lambda_V=0.00$). Similarly to the scaling results
discussed for $^{208}$Pb and $^{266}$Pb, this behaviour of $E_{M}^{\rm
scal}$ is a consequence of the fact that the matter radius of the
nucleus is smaller and the scaling incompressibility $K_A^{\rm scal}$
larger for a soft symmetry energy.

At variance with the uniformity displayed by the $E_{M}^{\rm scal}$
results, the constrained estimate $E_{M}^{\rm cons}$ of the GMR
excitation energy has a distinct feature with the symmetry energy
below and above $A\simeq 254$. If the mass number is below $A\simeq
254$, $E_{M}^{\rm cons}$ is larger when the parameter set has a soft
symmetry energy. However, this tendency is reversed from $A\simeq 254$
to the neutron drip line when the isotopes are more neutron rich.
Actually, in a given nucleus the $\Lambda_V$ parameter reduces its
neutron rms radius, which means that the system becomes on average
more compact. As long as the nucleus does not be very neutron rich, if
it is more compact, the effect of the constraining parameter $\eta$ on
the nuclear rms radius is less, and therefore the absolute value of
$(\partial R_{\eta}^2 / \partial \eta)_{\eta=0}$ in the denominator of
equation~(\ref{emcons2}) for $E_{M}^{\rm cons}$ is smaller. This
explains the larger magnitude of $E_{M}^{\rm cons}$ at
$\Lambda_V=0.03$, compared to $\Lambda_V=0.00$, in the Pb isotopes
lighter than $A\simeq 254$. But a point is reached in mass number
($A\simeq 254$ here) where the isotopes are so neutron rich that the
last neutrons become very weakly bound. These outer neutrons are very
soft against compression. As a result of their specific contribution
to $R_{\eta}^2$ when one applies an external constraint, the absolute
value of $(\partial R_{\eta}^2 / \partial \eta)_{\eta=0}$ increases
sharply instead of decreasing. This originates the large reduction of
$E_{M}^{\rm cons}$ seen in figure~\ref{f_pb_ex} for the more
neutron-rich isotopes. The reduction of $E_{M}^{\rm cons}$ at large
neutron numbers is enhanced when the symmetry energy is soft compared
to the case of a stiff symmetry energy because the outer neutrons are
more weakly bound in the former case. A comparable decrease of the
excitation energy of the GMR, however, is not present in the scaling
calculations because the discussed effect does not arise in a
self-similar scaling of the density of the nucleus. As we mentioned in
section \ref{resemblance}, the scaling and constrained average
energies pertain to different energy regimes of the GMR\@. While
$E_{M}^{\rm scal}$ informs more about the relatively high-energy
contributions to the RPA strength, $E_{M}^{\rm cons}$ carries more
information about the low-energy sector which is the region that shows
the more noticeable changes in largely neutron-rich nuclei because of
the presence of weakly bound neutrons.

We may take $\sigma= \frac{1}{2} [(E_{M}^{\rm scal})^2 - (E_{M}^{\rm
cons})^2]^{1/2}$ as an estimate of the width of the GMR, in analogy
with the nonrelativistic framework
\cite{bohigas79,lipparini89,gleissl90} (see section
\ref{resemblance}). From the separation observed in figure~\ref{f_pb_ex}
between $E_{M}^{\rm scal}$ and $E_{M}^{\rm cons}$,
at either $\Lambda_V=0.00$ or $\Lambda_V=0.03$, the width of
the resonance is expected to be almost independent of the mass number
and of the density dependence of the symmetry energy for all of the Pb
nuclei located in the region between the proton-rich side of the
isotopic chain till $A\simeq 230$--240. However, $E_{M}^{\rm cons}$
starts to show a pronounced departure from $E_{M}^{\rm scal}$ after
$A\simeq 240$ and, as a consequence, the resonance width should
display a sharp increase for the Pb isotopes with large neutron
numbers. We see from figure~\ref{f_pb_ex} that this effect is predicted
to be stronger when the nuclear interaction has a soft symmetry
energy. As discussed in \cite{cent05} in 
nonrelativistic calculations with Skyrme forces, the value of the
estimate $\sigma$ of the GMR width provides a qualitative idea about
the distribution of the RPA strength. A small width $\sigma$ indicates
that the RPA strength of the GMR is basically concentrated in a narrow
single peak, while a large width $\sigma$ suggests that the peak is
broad, or even that the RPA strength may be fragmented into several
peaks. The large value of $\overline{m}_{-1} \propto -(\partial
R_{\eta}^2 / \partial \eta)_{\eta=0}$ (see equation~(\ref{emcons2}))
that one finds in the constrained calculations for the more
neutron-rich Pb isotopes, points towards an enhancement of the RPA
strength in the low-energy region in these nuclei, due to transitions
from the last weakly bound single-particle levels to the continuum
\cite{cent05}. (This effect, which has a quantal origin, is accounted
at least on average by the semiclassical calculations~\cite{cent05}.)
The discussed scenario is actually confirmed by the QRPA calculations
of \cite{colo09} where the fragmentation as well as the enhancement of
the low-energy part of the RPA strength in Ca and Ni isotopes
approaching the neutron drip line can be seen in figures~3 and~6.

\begin{figure}
\includegraphics[width=0.70\columnwidth,angle=0,clip=true]
{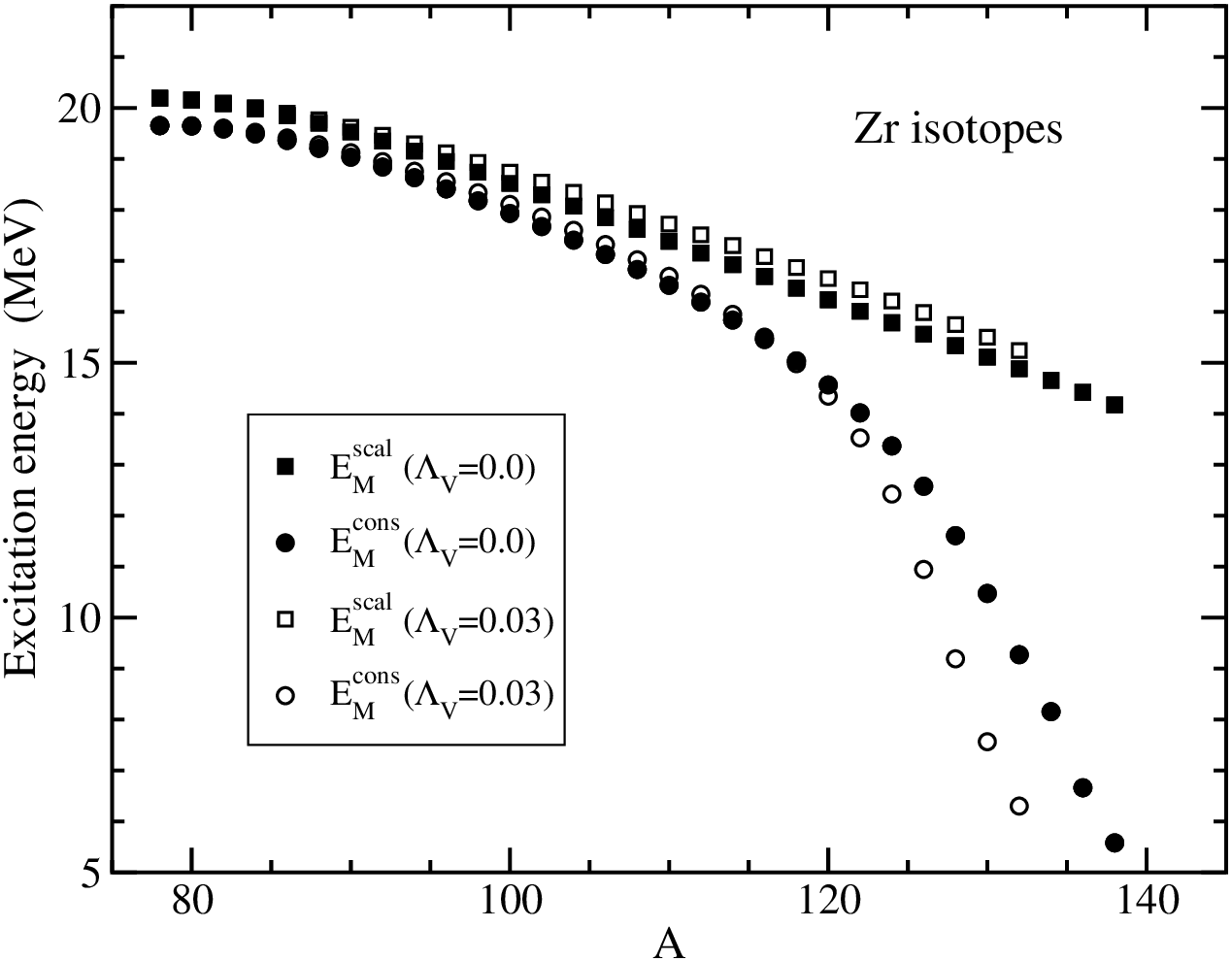}
\caption{Mass-number dependence of the excitation energy of the giant
monopole resonance in the Zr isotopic chain for the NL3 model with
$\Lambda_V=0$ (stiff symmetry energy) and $\Lambda_V=0.03$ (soft
symmetry energy), in the scaling and constrained approaches.}
\label{f_zr_ex}
\end{figure}

Lastly, we present in figure~\ref{f_zr_ex} the calculated results for
the excitation energies of the GMR in the Zr isotopic chain. The
variation of the excitation energies obtained in the scaling scheme as
a function of the mass number $A$ and of the $\Lambda_V$ coupling is
qualitatively similar to the pattern exhibited by the Pb isotopes.
That is, the excitation energies $E_{M}^{\rm scal}$ show a monotonic
decrease with increasing mass number, and they are larger when
calculated with a soft symmetry energy. The relative differences
between the values of $E_{M}^{\rm scal}$ at $\Lambda_V=0.03$ and at
$\Lambda_V=0.00$ are smaller compared to the chain of the Pb isotopes
because of the smaller isospin content of the Zr isotopes. Concerning
the results for the constrained excitation energies in Zr, the value
of $E_{M}^{\rm cons}$ shows a stronger dependence on the mass number
$A$ at large neutron numbers than the scaling value $E_{M}^{\rm
scal}$. As long as $A\lesssim 120$, the excitation energy $E_{M}^{\rm
cons}$ for zirconium is to a large extent independent of the symmetry
energy. But afterwards ($A\gtrsim 120$), $E_{M}^{\rm cons}$ becomes
visibly lower for the soft symmetry energy.

In consonance with the behaviour of $E_{M}^{\rm scal}$ and $E_{M}^{\rm
cons}$ found in figure~\ref{f_zr_ex}, it is predicted that the width
$\sigma$ of the GMR for the Zr nuclei should remain practically
constant till $A\simeq 100$. The width $\sigma$ would start increasing
moderately between $A\simeq 100$ and $A\simeq 120$, because of the
slight gradual separation between $E_{M}^{\rm scal}$ and $E_{M}^{\rm
cons}$ in these isotopes. Finally, the resonance width would be much
enhanced in the region of the most neutron-rich Zr isotopes, where the
effect is largely reinforced by having a soft symmetry energy in the
nuclear interaction.

\section{Summary and conclusions}

Though the excitation energy of the isoscalar giant monopole resonance
is mainly ruled by the compression modulus of symmetric nuclear
matter, it is also affected by the symmetry energy in neutron-rich
nuclei. In this paper we have analyzed the influence of the density
dependence of the symmetry energy on the average excitation energy of
the GMR\@. In order to do this study, we have used a relativistic mean
field model supplemented by an additional $\omega$-$\rho$ nonlinear
interaction driven by a coupling constant $\Lambda_V$. This additional
coupling allows one to generate different parameter sets that preserve
the properties of the symmetric matter and differ among them, only, in
the density dependence of the symmetry energy, which is softened by
increasing $\Lambda_V$. As a function of the coupling constant
$\Lambda_V$, the model predicts in finite nuclei almost the same
binding energy and proton rms radius as in the case where
$\Lambda_V=0$, while the neutron rms radius gets significantly reduced. 

The excitation energy of the isoscalar giant monopole resonance has
been computed using the semiclassical relativistic extended
Thomas-Fermi theory. We obtain the energy of the breathing mode in two
different ways: through the scaling model and performing constrained
calculations. It is known that these calculations with nonrelativistic
Skyrme forces nicely reproduce the average energies $E_3$ and $E_1$
obtained from the more microscopic RPA calculations, not only for
stable nuclei but also for largely neutron-rich nuclei \cite{cent05}.
In our study, we have calculated the scaling and constrained
semiclassical average energies with the relativistic model along the
isotopic chains of Pb and Zr. We can also obtain some qualitative
information about the distribution of the RPA strength from the
resonance width computed using these two semiclassical estimates of
the excitation energy.

As it is known, the excitation energy of the breathing mode decreases,
at least for stable nuclei, following roughly an $A^{-1/3}$ law
\cite{blai80} due to the geometrical enhancement of the mass
denominator. However, when the neutron-proton asymmetry increases,
departures from this behaviour appear. The reason is that the
isospin-dependent part of the finite nucleus incompressibility $K_{A}$
plays a non-negligible role by reducing the restoring force and making
nuclei with large neutron numbers softer than the stable isotopes.
This effect in the average excitation energy is present in the scaling
estimate but it is particularly strong in the constrained calculation.
When the neutron-proton asymmetry increases, the outer neutrons become
more and more weakly bound starting to behave almost as extremely
asymmetric nuclear matter, which is very soft compared to symmetric
matter \cite{Bom91}. This effect makes the compression modulus $K^{\rm
cons}_{\rm A}$ smaller and, therefore, the average excitation energy
of the GMR decreases. As discussed in the case of constrained HF
calculations with Skyrme forces \cite{cent05}, this scenario reflects
the important contribution to the low-energy part of the RPA strength
distribution coming from the particle-hole transitions to the
continuum from the weakly bound neutron energy levels in nuclei having
large neutron numbers. A similar situation has been found recently by
Capelli, Col\`o and Li \cite{colo09} in QRPA and constrained HFB
Skyrme calculations performed along the light isotopic chains of Ca
and~Ni. 

For a given nucleus the excitation energy of the GMR also depends on
the density content of the symmetry energy of the model used to
compute it. In stable nuclei the scaled and constrained estimates of
the excitation energy are larger for parameter sets that have a
symmetry energy with a soft density dependence. In the case of the
constrained estimate, this tendency is reversed in the nuclei with
large neutron-proton asymmetry. The reason lies in the fact that in
these nuclei the outermost and weakly bound neutrons are very
compressible and consequently the constrained compression modulus
$K_{\rm A}^{\rm cons}$ becomes lower when the density dependence of
the symmetry energy is soft (the outer neutrons are then less bound)
than when the symmetry energy is stiff. Thus, the average excitation
energies based on the inverse energy-weighted sum rule are predicted
to be significantly reduced for large neutron numbers when computed
with a parameter set that has a softer density dependence of the
symmetry energy.

In conclusion, the density dependence of the symmetry energy has a
moderate influence on the value of the excitation energy of the
isoscalar giant monopole resonance in stable nuclei and in nuclei that
are not far from stability. It has a larger impact on the GMR of
isotopes with large isospins, where the calculations reveal that the
constrained energies become much reduced in models having a soft
symmetry energy. The results point towards a considerable enhancement
of the low-energy part of the RPA strength and to likely different
structures in the excitation spectra of the GMR in unstable
neutron-rich nuclei if the nuclear symmetry energy is soft. It may be
worth pursuing a detailed characterization of these predictions
related with the softness of the symmetry energy by computing the
strength distribution in a few selected isotopes at large neutron
numbers with the more demanding but more microscopic QRPA theory of
the GMR response. Unfortunately, it is difficult that the indicated
effects be tested in the laboratory as they become prominent at
neutron numbers beyond present reach. Such nuclear systems are
otherwise expected to exist in astrophysical conditions, such as those
prevailing in the inner crust of neutron stars, where exotic nuclear
clusters can be stabilized by an enveloping gas of neutrons and
electrons \cite{sil02}. It is thought that low-energy collective
excitations play an important role in astrophysical processes that
involve nuclei far from stability \cite{paar07}, and in that context
the modification of the low-energy region of the strength distribution
of the monopole resonance by the softness of the symmetry energy could
have implications.

\begin{acknowledgments}
M.C., X.R. and X.V. would like to thank J. Piekarewicz for useful
conversations. They acknowledge support from the Consolider Ingenio
2010 Programme CPAN CSD2007-00042 and grants FIS2008-01661 from MEC
and FEDER, and 2009SGR-1289 from Generalitat de Catalunya. X.R. also
acknowledges grant AP2005-4751 from MEC\@. S.K.P. and B.K.S.
acknowledge the support of CSIR (No.\ 03(1060) 06/EMR-II), Government
of India. P.D.S. acknowledges support from the UK Science and Technology 
Facilities council under grants PP/F000596/1 and ST/F012012/1.
\end{acknowledgments}

\newpage

%
\appendix

\section{Equations of the relativistic extended Thomas-Fermi model}

We have introduced the energy density of the relativistic nuclear mean
field model supplemented by an isoscalar-isovector mixed
$\omega$-$\rho$ interaction in equation~(\ref{eq0a}) of the text.
Following 
\cite{centelles93a,centelles92,speicher93,centelles98}, 
in the relativistic extended Thomas-Fermi approach the expression of
the nucleonic part ${\cal E}$ is written as ${\cal E}={\cal E}_0+{\cal
E}_2$, where
\begin{equation}
{\cal E}_0 =
\sum_{q={\rm n,p}} \frac{1}{8\pi^2} \left[k_{{\rm
F}q}\epsilon^{3}_{{\rm F}q} +k^{3}_{{\rm F}q}\epsilon_{{\rm F}q}
-{m^*}^{4}\ln\frac{k_{{\rm F}q}+\epsilon_{{\rm F}q}}{m^*}\right]
\label{eq0c}
\end{equation}
and
\begin{equation}
{\cal E}_2 =  \sum_{q={\rm n,p}} \left[ C_{1q}(k_{{\rm F}q}, m^*)
 (\vect{\nabla} \rho_{q})^2
+ C_{2q} (k_{{\rm F}q}, m^*)\left( \vect{\nabla}
\rho_{q} \cdot \vect{\nabla} m^* \right)
+ C_{3q}(k_{{\rm F}q}, m^*)
(\vect{\nabla} m^*)^2 \right] .
\label{eq0d}
\end{equation}
The leading term ${\cal E}_0$ is the contribution of the pure
Thomas-Fermi part, while the term ${\cal E}_2$ contains the gradient
corrections of order $\hbar^2$. The variables $k_{{\rm F}q}$, $m^*$,
and $\epsilon_{{\rm F}q}$ are, respectively, the local Fermi momentum
$k_{{\rm F}q}= (3\pi^2 \rho_{q})^{1/3}$, the Dirac effective mass
$m^*=m-\Phi$, and the dispersion relation $\epsilon_{{\rm
F}q}=(k^2_{{\rm F}q}+{m^*}^2)^{1/2}$ of the effective particle, with
$q=\rm n$ for neutrons and $q=\rm p$ for protons. The coefficients
$C_{iq}$ stand for the following analytic functions of $k_{{\rm F}q}$
and $m^*=m-\Phi$:
\begin{eqnarray}
C_{1q} & = & \frac{\pi^2}{24 k_{{\rm F}q}^3
\epsilon_{{\rm F}q}^2}
\left( \epsilon_{{\rm F}q} + 2k_{{\rm F}q} \ln \frac{k_{{\rm F}q} +
\epsilon_{{\rm F}q}}{m^*} \right) ,
\nonumber \\[3mm]
C_{2q} & = &
\frac{m^*}{6 k_{{\rm F}q} \epsilon_{{\rm F}q}^2}
\ln \frac{k_{{\rm F}q} + \epsilon_{{\rm F}q}}{m^*} ,
\nonumber\\[3mm]
C_{3q} & = &
\frac{k_{{\rm F}q}^2}{24 \pi^2 \epsilon_{{\rm F}q}^2}
\left[\frac {\epsilon_{{\rm F}q}}{k_{{\rm F}q}}
- \left( 2 + \frac{\epsilon_{{\rm F}q}^2}{k_{{\rm F}q}^2} \right)
\ln \frac{k_{{\rm F}q} + \epsilon_{{\rm F}q}}{m^*} \right] .
\label{eq0f}
\end{eqnarray}

The RETF ground-state densities and meson fields are calculated by
solving by numerical iteration the Euler--Lagrange equations
associated to the energy density (\ref{eq0a}), with the constraint of
baryon number conservation (which is imposed through Lagrange
multipliers $\mu_q$, the chemical potentials of neutrons and protons):
\begin{eqnarray}
& & \epsilon_{{\rm F}q} + W 
    + \frac{1}{2} [{\cal A} + ({\cal A}+B)\tau_{3q}]
\nonumber \\[3mm]
& &
 - 2 C_{1q}\Delta\rho_{q} - C_{2q}\Delta m^*
- \frac{\partial C_{1q}}{\partial \rho_{q}}(\vect{\nabla} \rho_q)^2 
- 2 \frac{\partial C_{1q}}{\partial m^*}
(\vect{\nabla} \rho_{q} \cdot \vect{\nabla} m^*)
\nonumber \\[3mm]
& & 
-  \left( \frac{\partial C_{2q}}{\partial m^*}
- \frac{\partial C_{3q}}{\partial \rho_{q}} \right)
(\vect{\nabla} m^* )^2 = m + \mu_q \,,
\label{eq0h}
\end{eqnarray}
\begin{eqnarray}
(\Delta- m_{s}^2)\Phi & = & -g_{s}^2 \left( \rho_{s} 
- \frac{\kappa}{2!} \Phi^2 - \frac{\lambda}{3!} \Phi^3 \right) 
\;\equiv\; -g_{s}^2 \rho^{\rm eff}_{s} ,
\label{eqFN4}  \\[3mm]
   (\Delta - m_{v}^2) W  & = &   -g_{v}^2 \left( \rho 
-  2 \Lambda_V B^2 W \right) ,
\label{eqFN5}  \\[3mm]
   (\Delta - m_\rho^2) B  & = &  - g_\rho^2 \left( \rho_3 
- 2 \Lambda_V B W^2 \right) ,
\label{eqFN6}  \\[3mm]
 \Delta {\cal A}   & = &   - \rho_{p} \,.
\label{eqFN7}
\end{eqnarray}
We have $\tau_{3q}=+1$ for protons and $\tau_{3q}=-1$ for neutrons.
The baryon density is $\rho= \rho_{p}+\rho_{n}$, while $\rho_3=
{\textstyle\frac{1}{2}} (\rho_{p}-\rho_{n})$ is the isovector density
of the nucleus. Finally, the RETF expression of the scalar density
$\rho_{s}$ entering equation~(\ref{eqFN4}) is given by
\begin{eqnarray}
\rho_{s} & = & \frac{\delta {\cal E}}{\delta m^*}
\;=\; \frac{\delta {\cal E}_0}{\delta m^*}
+ \frac{\delta {\cal E}_2}{\delta m^*}
\;=\; \rho_{s0} + \rho_{s2}
\nonumber \\[3mm]
& = &
\sum_{q} \frac{m^*}{2\pi^2}\left[k_{{\rm F}q}\epsilon_{{\rm
F}q}-{m^*}^2 \ln\frac{k_{{\rm F}q}
+\epsilon_{{\rm F}q}} {m^*}\right]
\nonumber \\[3mm]
& &
 - \sum_q \left[ C_{2q}\Delta\rho_{q}+2C_{3q}\Delta m^*
+\left(\frac{\partial C_{2q}}
{\partial \rho_{q}}-\frac{\partial C_{1q}}{\partial m^*}\right)
 (\vect{\nabla} \rho_q)^2 \right.
\nonumber \\[3mm]
& & \left. \mbox{}
+2 \frac{\partial C_{3q}}{\partial \rho_{q}}
(\vect{\nabla} \rho_{q} \cdot \vect{\nabla} m^*)
 + \frac{\partial C_{3q}}{\partial m^*}
(\vect{\nabla} m^* )^2 \right] .
\label{eq0g}
\end{eqnarray}

\section{Klein-Gordon equations of the scaled meson fields}

In this appendix we illustrate how to calculate the derivatives of the
scaled meson fields upon the scaling parameter $\alpha$ that we have
used to represent the collective coordinate of the monopole
oscillation of the nucleus in the scaling approach. In view of
equation~(\ref{eq4}) for the scaled baryon density $\rho_\alpha$,
i.e., $\rho_\alpha (\vect{r}) = \alpha^3 \rho(\alpha\vect{r})$, and of
the field equation (\ref{eqFN5}) obeyed by the omega-meson field $W$,
the scaled field $W_\alpha$ fulfills the Klein--Gordon equation
\begin{equation}
\left( \Delta_{\vect{\scriptstyle u}}
- \frac{m_{v}^2}{\alpha^2} \right) W_\alpha(\vect{u}) =
- g_{v}^2 \left( \alpha \,\rho(\vect{u})
- \frac{2 \Lambda_V}{\alpha^2}B_\alpha(\vect{u})^2 
W_\alpha(\vect{u}) \right) , 
\label{eqap1}
\end{equation}
where we have introduced the coordinate $\vect{u}\equiv \alpha
\vect{r}$. On differentiating equation~(\ref{eqap1}) with respect to the
scaling parameter $\alpha$ we have
\begin{eqnarray}
& & \left( \Delta_{\vect{\scriptstyle u}}
- \frac{m_{\rm v}^2}{\alpha^2} \right) 
\frac{\partial W_\alpha}{\partial\alpha} =
\nonumber \\[3mm]
& & -g_{v}^2 \left( \rho 
+ \frac{2 m_{v}^2}{g_v^2}\frac{W_\alpha}{ \alpha^3}
+ \frac{4 \Lambda_V}{\alpha^3}B_\alpha^2 W_\alpha
- \frac{4 \Lambda_V}{\alpha^2}B_\alpha W_\alpha
\frac{\partial B_\alpha}{\partial\alpha}
- \frac{2 \Lambda_V}{\alpha^2}B_\alpha^2 
\frac{\partial W_\alpha}{\partial\alpha} \right) .
\label{eqap2}
\end{eqnarray}
If we now set $\alpha= 1$, the solution of this equation allows us to
obtain the value of the derivative $\partial W_\alpha /\partial\alpha
|_{\alpha=1}$. One can work out the equations for the scaled isovector
$\rho$-meson field in the same way. In practice, one simply has to
replace ($W_\alpha,\, B_\alpha,\, \rho,\, m_v,\, g_v$) by
($B_\alpha,\, W_\alpha,\, \rho_3,\, m_\rho,\, g_\rho$) in
equations~(\ref{eqap1}) and (\ref{eqap2}).

In the case of the scalar field one has to take account of the terms
that arise due to the fact that the scalar density $\rho_{s}$ is
itself a function of the scalar field. Following the same steps as
above, from the Klein-Gordon equation
\begin{equation}
\left( \Delta_{\vect{\scriptstyle u}}
- \frac{m_{\rm s}^2}{\alpha^2} \right) \Phi_\alpha(\vect{u}) =
- \alpha g_{s}^2 {\tilde\rho}^{\rm eff}_{s}(\vect{u} )
\label{eqap6}
\end{equation}
for the scaled scalar field $\Phi_\alpha$, one readily arrives at 
the equation
\begin{equation}
\left( \Delta_{\vect{\scriptstyle u}}
- \frac{m_{s}^2}{\alpha^2} \right)
\frac{\partial \Phi_\alpha}{\partial\alpha} =
- g_{\rm s}^2 \left( {\tilde\rho}^{\rm eff}_{\rm s} + \alpha
\frac{\partial {\tilde\rho}^{\rm eff}_{s}}{\partial \alpha}
+ \frac{2 m_{s}^2}{g_s^2} \frac{\Phi_\alpha}{\alpha^3} \right),
\label{eqap4}
\end{equation}
whose solution at $\alpha=1$ provides the value of $\partial
\Phi_\alpha / \partial\alpha |_{\alpha=1}$.

%

%
\end{document}